\def \SAIT #1 #2 #3 {{\em Mem.\ Soc.\ Astron.\ It.\/}  {\bf #1}  #2 #3}  
\def \MESS #1 #2 {{\em The Messenger\/} {\bf #1}, #2}
\def \ASTRNACH #1 #2 {{\em Astron. Nach.\/} {\bf #1}, #2}
\def \AAP #1 #2 {{\em Astron. Astrophys.\/} {\bf #1}, #2}
\def \AAL #1 #2 {{\em Astron. Astrophys. Lett.\/} {\bf #1}, L#2}
\def \AAR #1 #2 {{\em Astron. Astrophys. Rev.\/} {\bf #1}, #2}
\def \AAS #1 #2 #3  {{\em Astron. Astrophys. Suppl. Ser.\/} {\bf #1}  #2 #3}   
\def \AJ #1 #2 {{\em Astron. J.\/} {\bf #1}, #2}
\def \ANNREV #1 #2 {{\em Ann. Rev. Astron. Astrophys.\/} {\bf #1}, #2}
\def \APJ #1 #2 #3 {{\em Astophys. Journal\/} {\bf #1}  #2  #3}   
\def \APJL #1 #2 {{\em Astrophys.. J. Lett.\/} {\bf #1}, L#2}
\def \APJS #1 #2 {{\em Astrophys. J. Suppl.\/} {\bf #1}, #2}
\def \APSS #1 #2 {{\em Astrophys. Space Sci.\/} {\bf #1}, #2}
\def \ASR #1 #2 {{\em Adv. Space Res.\/} {\bf #1}, #2}
\def \BAIC #1 #2 {{\em Bull. Astron. Inst. Czechosl.\/} {\bf #1}, #2}
\def \JSQRT #1 #2 {{\em J. Quant. Spectrosc. Radiat. Transfer\/} {\bf #1},#2}
\def \MN #1 #2 {{\em Mon. Not. R. Astr. Soc.\/} {\bf #1}, #2}
\def \MEM #1 #2 {{\em Mem. R. Astr. Soc.\/} {\bf #1}, #2}
\def \PLR #1 #2 {{\em Phys. Lett. Rev.\/} {\bf #1}, #2}
\def \PASJ #1 #2 {{\em Publ. Astron. Soc. Japan\/} {\bf #1}, #2}
\def \PASP #1 #2 {{\em Publ. Astr. Soc. Pacific\/} {\bf #1}, #2}
\def \NAT #1 #2 {{\em Nature\/} {\bf #1}, #2}

\newcommand{\be}{\begin{equation}}
\newcommand{\ee}{\end{equation}}

\documentclass[12pt]{iopart}
\usepackage{epsfig} 
\begin{document}

\title[ANTIMATTER RESEARCH IN SPACE]{ANTIMATTER RESEARCH IN SPACE}

\author{Piergiorgio Picozza  and Aldo Morselli}

\address{University of Roma "Tor Vergata" and INFN Roma 2, Roma, Italy }

\begin{verse} 
 {\small    \hspace{6cm}  We must regard it rather an accident 
\\   \hspace{6cm}  that the Earth and presumably the whole 
\\  \hspace{6cm}   Solar System contains a preponderance 
\\  \hspace{6cm}   of negative electrons and positive protons.
\\  \hspace{6cm}   It is quite possible that for some of the stars  
\\  \hspace{6cm}   it is the other way about. 
         \vspace{0.4cm}
         \\  \hspace{9cm}  {\sl Paul Dirac in his speech 
         \\ \hspace{9cm}         accepting the Nobel Prize 
         \\ \hspace{9cm}         in Physics, (1933) }    }

 \end{verse}

\begin{abstract}
Two of the most compelling issues facing
astrophysics and cosmology today are to understand
the nature of the dark matter that pervades the
universe  and to understand the apparent
absence of cosmological antimatter. For both
issues, sensitive measurements of cosmic-ray
antiprotons and positrons, in a wide
energy range, are crucial. 

Many different mechanisms can contribute to antiprotons and positrons 
production, ranging from conventional reactions like 
$p+p\rightarrow p+\overline p+${\sl anything} up to exotic processes like 
 neutralino annihilation. The open problems are so fundamental 
 (i.e.: is the universe symmetric in matter and antimatter ?)
 that experiments in this field will probably be of the greatest 
 interest in the next years.
 
Here we will summarize the present situation, 
 showing the different hypothesis and models and the experimental 
 measurements needed to lead to a more established scenario.\end{abstract}



\maketitle
\bigskip

\section{Introduction}
The idea of exploiting cosmic antiprotons measurements to probe
unconventional particle physics and astrophysics scenarios has a long history [1-10]
and moved the cosmologists for several decades. Shortly after the discovery of the CP
violation in the weak interactions in '64, Sakharov formulated his famous hypotheses that were assumed
to be a reasonable starting point to explain the apparent contradiction between the fundamental
laws of the nature and the observations. Several balloon borne experiments were dedicated to the
search for antiparticles and antinuclei, and in the 70's the teams of R.~Golden in USA \cite{golden} and of
E.~Bogomolov in Russia \cite{bogomolov} identified the first antiprotons in cosmic rays (the positrons were
discovered more than 40 years before by Anderson). Then, the antiproton spectrum was intensively
studied for searching for signals exceeding the background of the antiprotons produced in the
interactions of CR's with the interstellar matter.

\begin{figure}
  \begin{center}
    \mbox{\epsfig{file=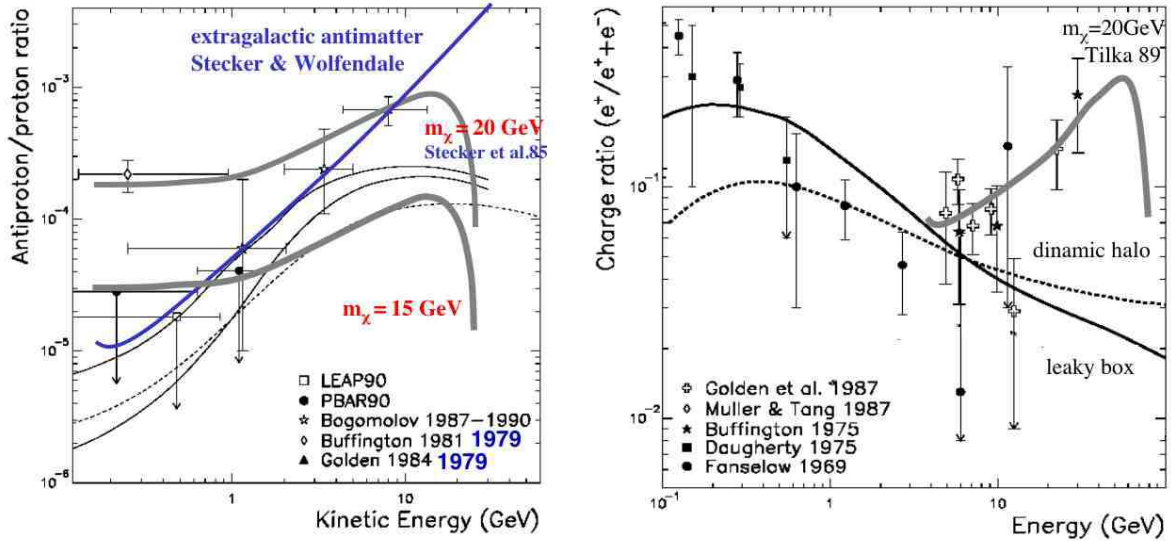,height=8.cm} }
  \end{center}
 \caption{Antiproton/proton ratio experimental situation before 1990 (on the left). For the experimental data the references are: LEAP \cite{leap}, PBAR~\cite{pbar},  E.~Bogomolov et al.~ \cite{bogomolov}, Buffington et al. \cite{buffington}, Golden et al.~\cite{golden}. Positron fraction experimental situation before 1990 ( on the right) compared
 with the foreseen secondary production in the interstellar medium and an exotic model. For the experimental data the references are: Golden et al.~\cite{golden87e},    
M\"uller \& Tang~\cite{muller_e},  Buffington et al. \cite{buffington_e},  Daugherty et al.~\cite{Daugherty}, Fanselow et al.~\cite{Fanselow69_e} }
\label{beaf90}
\end{figure}

In figure~\ref{beaf90} (on the left) are shown the data on the antiproton/proton ratio 
before 1990. At that time the standard production models  (black lines) \cite{mod} could not account for all the antiprotons measured by the Golden et al.  and Buffington et al. \cite{buffington} experiments, and this triggered the formulation of many exotic models 
ranging from antiprotons coming from antigalaxies \cite{stecker85}  to annihilation of supersimmetric dark matter (gray curve) \cite{stecker85s}.

In the same years the results of the positron ratio measurements were somewhat similar, with the experiments giving an higher flux of positron at energies greater than 10 GeV (see figure~\ref{beaf90} on the right), explained only with some exotic production, like again the annihilation of WIMPs  giving a contribution as shown by the  gray curve in the figure~\ref{beaf90} (from \cite{tilka}).

\section{The second generation experiments }

In the last decade  there has been an increase of experiments performed
using novel techniques developed for accelerator physics. This has permitted to improve
the statistical and systematic significance of the experimental measurements.
The WiZard Collaboration did several balloon flights, MASS89\cite{GoldenMPWC}, MASS91\cite{bas99}, TS93\cite{Bocciolini}, CAPRICE94\cite{cap94}  and CAPRICE98\cite{cap98},  exploring  an energy interval from some hundreds MeV to 50 GeV of the antiproton and positron spectra.

The core parts of the instrument were a magnetic spectrometer composed by  a superconducting magnet and a tracking system for charge sign and momentum measurements, an imaging calorimeter (streamer tube or silicon tungsten) and a $\beta$ selector (Cerenkov, TRD or RICH).
 
 \begin{figure}
  \begin{center}
    \mbox{\epsfig{file=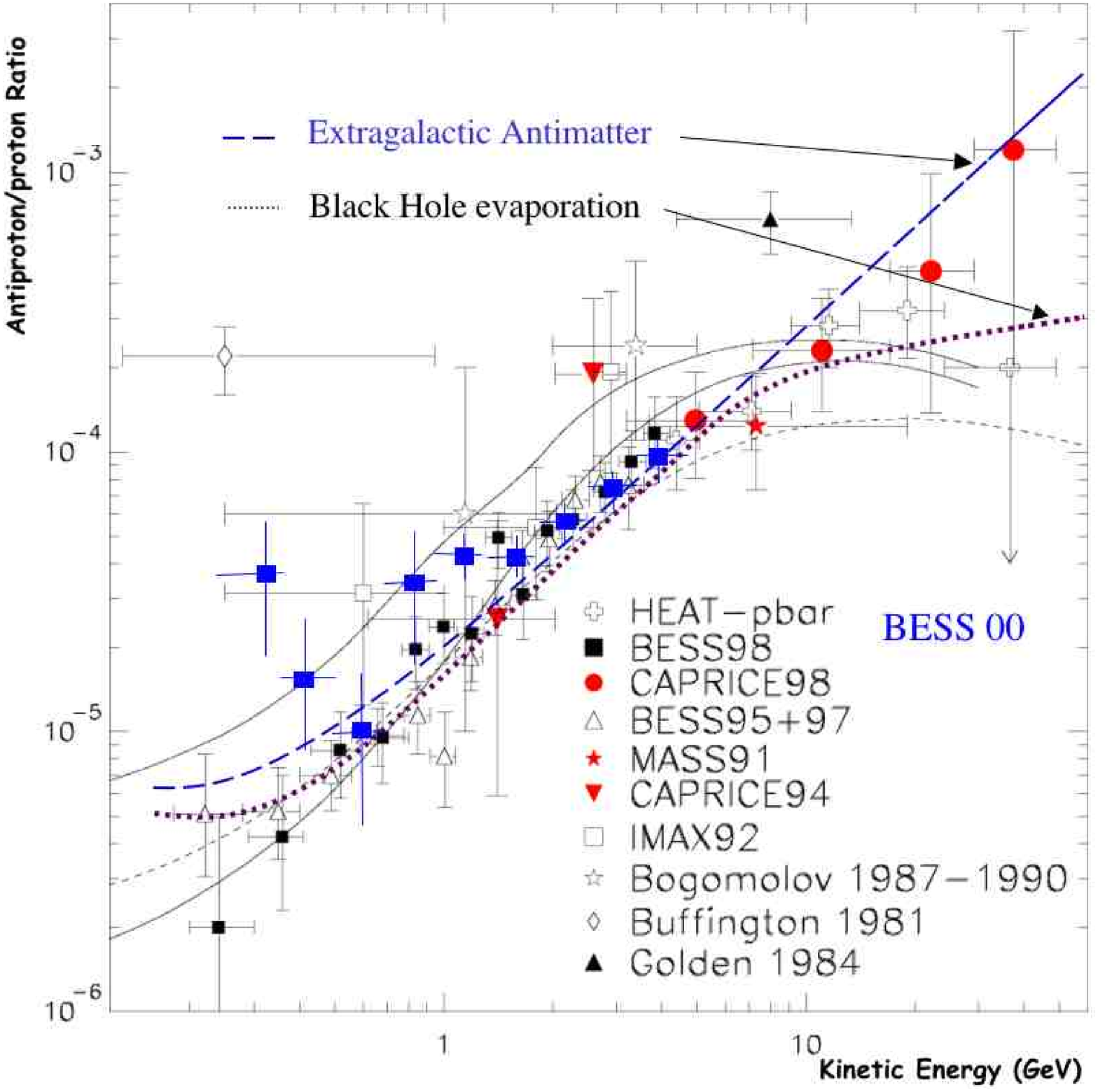,height=9.5cm} }
  \end{center}
    \caption {Antiproton proton ratio: experimental data and theoretical predictions. For the experimental data the references are: BESS~\cite{bess98},  Heat-pbar~\cite{bar98}, CAPRICE98~\cite{antip_data}, MASS91~\cite{bas99}, CAPRICE94~\cite{cap94antip}, 
IMAX92~\cite{mit96}, Bogomolov et al.~\cite{bogomolov}, Buffington et al.~\cite{buffington}, 
Golden et al.~\cite{golden}  }
    \label{antip}
\end{figure} 
 \begin{figure}
  \begin{center}
    \mbox{\epsfig{file=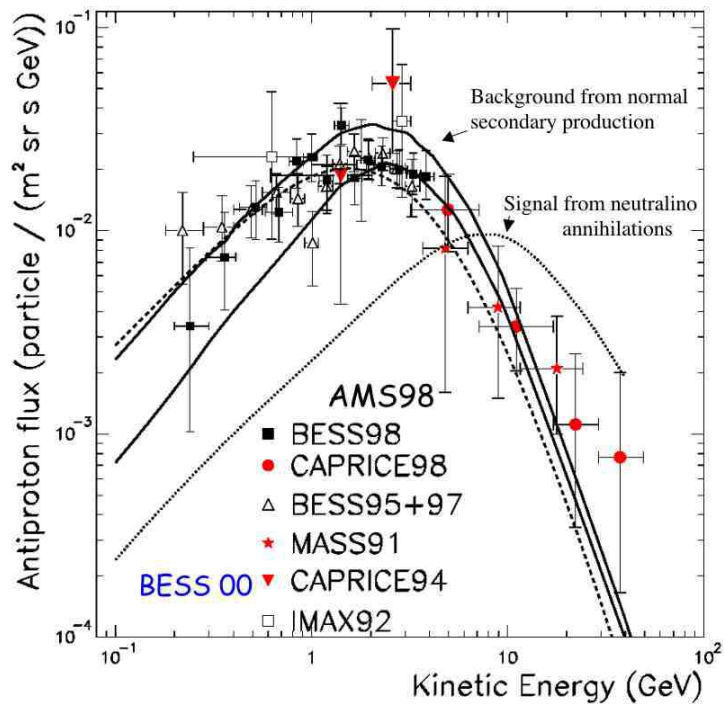,height=9.5cm} }
  \end{center}
    \caption {Antiproton absolute flux:  experimental data and theoretical predictions }
    \label{antipf}
\end{figure} 
 \begin{figure}
  \begin{center}
    \mbox{\epsfig{file=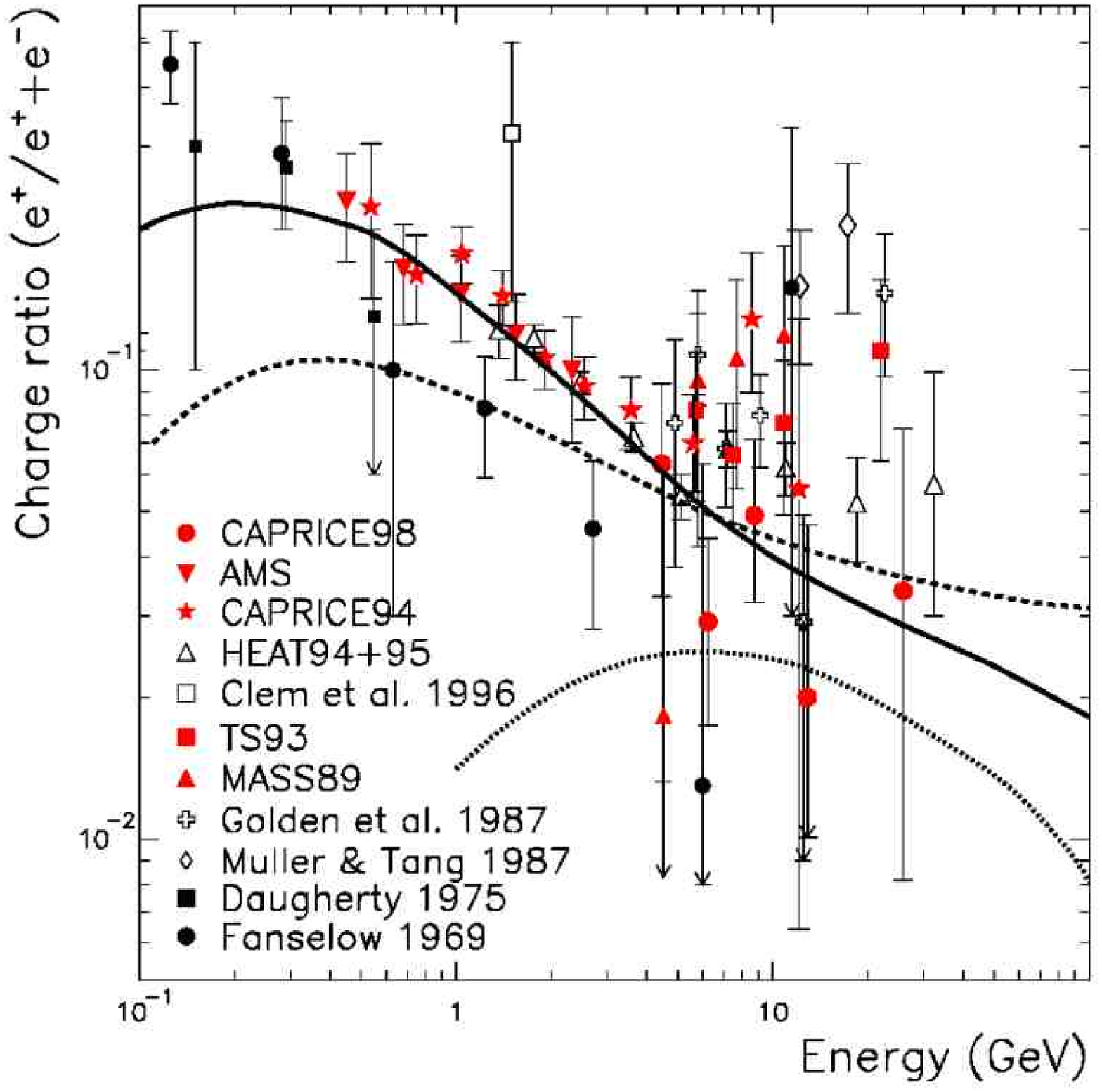,height=9.5cm} }
  \end{center}
    \caption {Positron fraction:  experimental data and theoretical predictions.  For the experimental results the references are: CAPRICE98~\cite{cap98e}, AMS~\cite{AMSe},  CAPRICE94~\cite{cap94e}, HEAT\cite{bar98}, Clem et al.~\cite{clem}, TS93~\cite{ts93e},    MASS89~\cite{mass89e},
Golden et al.~\cite{golden87e}, M\"uller \& Tang~\cite{muller_e}, Fanselow et al.~\cite{Fanselow69_e}   }
    \label{pos}
\end{figure} 

In  figure~\ref{antip}
there are reported the  data   for the antiproton/proton ratio collected up to now, together with the data of
HEAT-pbar \cite{bea01},  HEAT\cite{bar98}   and IMAX92\cite{mit96} experiments. 

The solid and dashed lines are,
respectively, the theoretical interstellar and solar modulated
 ratios for a pure secondary origin of the antiprotons.
 The others are more exotic models.

In  figure~\ref{antipf}
there are the  experimental results, including the AMS data\cite{AMSp},    for the antiproton flux  along with different theoretical 
calculations which account for a pure secondary component\cite{sim98,ber99c}
and together with the 
distortion on the antiproton flux (dashed line) due to  a possible
contribution from neutralino annihilation (dotted line, from\cite{Ullio}).

In figure~\ref{pos}  there are the experimental data for  the  $e^+ / (e^+ +e^- )$
fraction  together with the distortion  (solid line) due to
a possible contribution from neutralino annihilation (dotted line, from\cite{pam_e}).

The possibility to detect this kind of distortions and then the possibility to discover the nature of the dark matter 
is not so exotic as it might appear. 

Over the last   years our knowledge 
of the inventory of matter and energy in the universe has improved 
dramatically.  Astrophysical measurements from several experiments 
are now converging and a standard cosmological model is emerging.
For this model  the total matter density is about
40\%$\pm$10\% of the critical density of the Universe, with a contribution of the baryonic dark matter  less than
5\% and a contribution from neutrinos that cannot be greater than 5\%.  The remaining matter should be 
 composed of yet-undiscovered Weakly Interacting Massive Particles (WIMP) , and a good candidate for WIMP's 
is the Lightest Supersymmetric Particle (LSP) in R-parity conserving supersymmetric  models.  

 The best candidate as LSPs are  the Neutralinos. They are Majorana fermions and annihilate with each other
in the galactic halo producing leptons, quarks, gluons, gauge bosons and Higgs
bosons. These decay and/or
form jets that will give rise to antiprotons (and antineutrons which
decay shortly to antiprotons), positrons and gammas. A description of the signature of the annihilation in gamma spectra can be found in~\cite{Morselli_trento}.  
From the figures ~\ref{antipf} and~\ref{pos} it is clear that new experiments with better statistic are needed to discriminate among the models.

\begin{figure}
  \begin{center}
    \mbox{\epsfig{file=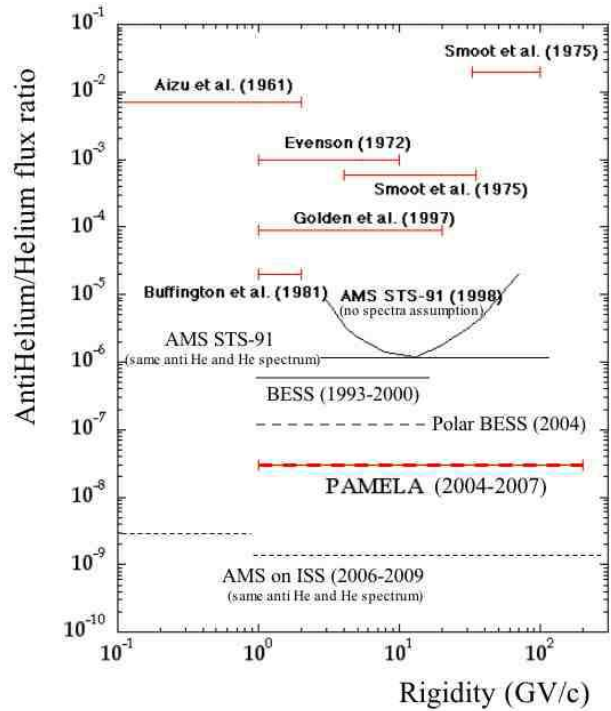,height=9.5cm} }
  \end{center}
    \caption {Present experimental limits  for the antihelium/helium ratio   and the sensitivities foreseen for the 
next experiments }
    \label{antinu}
\end{figure} 

An other goal of these experimens is the search of antinuclei.
Detection of heavy ( Z$>$2) antimatter of primary origin in cosmic rays would be a discovery of fundamental
significance.  That  would provide direct evidence of the existence of antimatter in the
universe.  The current theory suggests that the remaining
matter is the remnant of the almost complete annihilation of matter and antimatter at some early
epoch, which stopped only when there was no more antimatter to annihilate. Starting from a
matter/antimatter symmetric Universe, the required conditions for a following asymmetric evolution
are the CP violation, the baryon number non-conservation and a non equilibrium environment. On the
basis of gamma-ray observations, the coexistence of condensed matter and antimatter on scales
smaller than that of clusters of galaxies has been virtually ruled out. However, no observations
presently exclude the possibility that the domain size for establishing the sign of CP violation is
as large as a cluster or super-cluster of galaxies. For example, there could be equality in the
number of super-clusters and anti-super-clusters. Similarly, there is nothing that excludes the
possibility that a small fraction of the cosmic rays observed at Earth reaching our Galaxy from nearby
super-clusters.  Up to now no antihelium or heavier antinucleus have been found; 
in figure~\ref{antinu}  there are the present experimental limits     and the sensitivities foreseen for the 
next experiments that will be discussed in the next paragraphs.

\section{The PAMELA mission}

Three major experiments are being prepared for the future :  the  polar balloon  BESS instrument, 
the satellite PAMELA experiment and AMS on the International Space Station.

 \begin{figure}
  \begin{center}
    \mbox{\epsfig{file=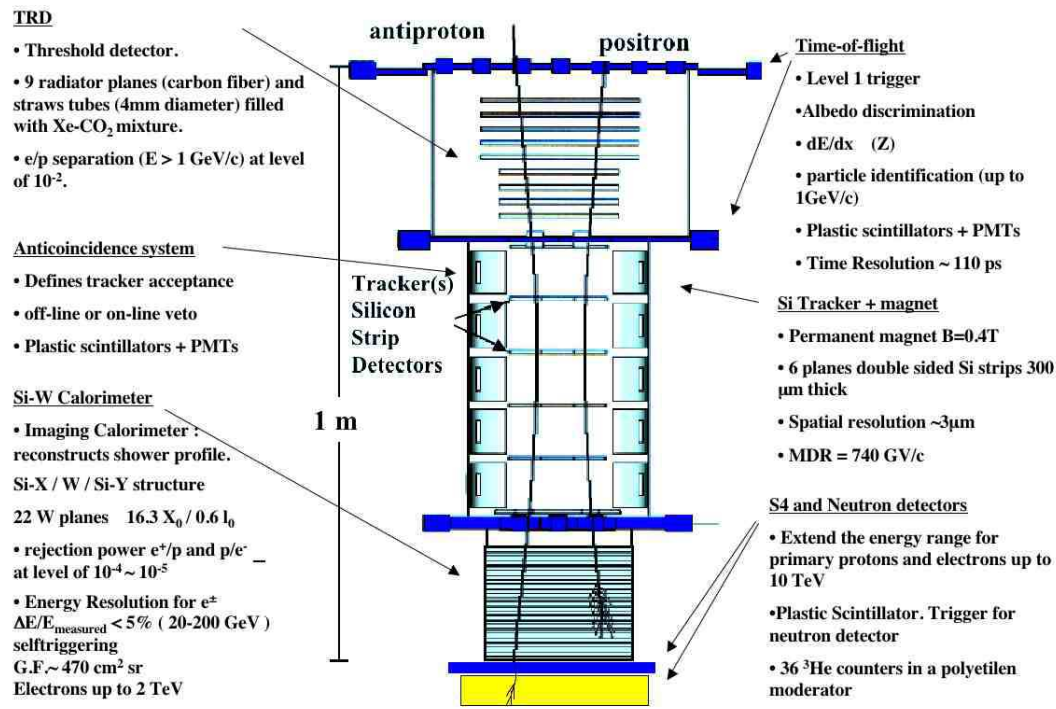,height=9.5cm} }
  \end{center}
    \caption {Schematic of the PAMELA  baseline instrument }
    \label{pamd}
\end{figure} 

 The PAMELA telescope,  shown in figure~\ref{pamd},  is a satellite-borne apparatus  built by the WiZard-PAMELA collaboration
\cite{pamela}.  The list of the people and  the Institution involved in the collaboration together with the on-line
status of the project is available at {\sl http://wizard.roma2.infn.it/}.
\subsection {PAMELA Scientific Objectives}

The PAMELA scientific primary goal is the search for heavy nuclei and non baryonic particles outside the Standard Model,
for the understanding of the formation and evolution of our Galaxy and the Universe and for the exploring of the cycles of matter and energy in the Universe.
Additional objectives of PAMELA are the studies of galactic cosmic rays in the heliosphere, Solar flares, 
distribution and acceleration of solar cosmic rays (SCR's) in the internal heliosphere, magnetosphere and magnetic field of the Earth, stationary and disturbed fluxes of high energy particles in the Earth's  magnetosphere and  anomalous components of cosmic rays.

\vskip 0.2cm
The PAMELA observations will extend the results of balloon-borne experiments over an
unexplored range of energies with unprecedented statistics and will complement the
information gathered from the Great Space Observatories.

More precisely, during its three years of planned operation, PAMELA will measure with very high statistics:

$\bullet$ Positron	flux from 50 MeV to  270 GeV ~~(present limits 0.7 - 30 GeV)

 $\bullet$ Antiproton flux	from  80 MeV to 190 GeV ~~ (present limits 0.4 - 50 GeV )

 $\bullet$ Limit on antinuclei $\sim10^{-8} ({\overline{He}} /He)$   ~~ (present limit about $10^{-6}$)

$\bullet$ Electron	flux	from 50 MeV to 3TeV

 $\bullet$ Proton	flux	from 80 MeV to 700 GeV

$\bullet$ Light  nuclei flux (up to oxygen) from 100 MeV/n  to 200 GeV/n  

$\bullet$ Electron and proton components up  to 10 TeV

In addition it will assure a continuous monitoring of the cosmic rays solar modulation 	

\vskip 0.2cm

\begin{figure}
\vspace{0cm}
    \epsfig{file=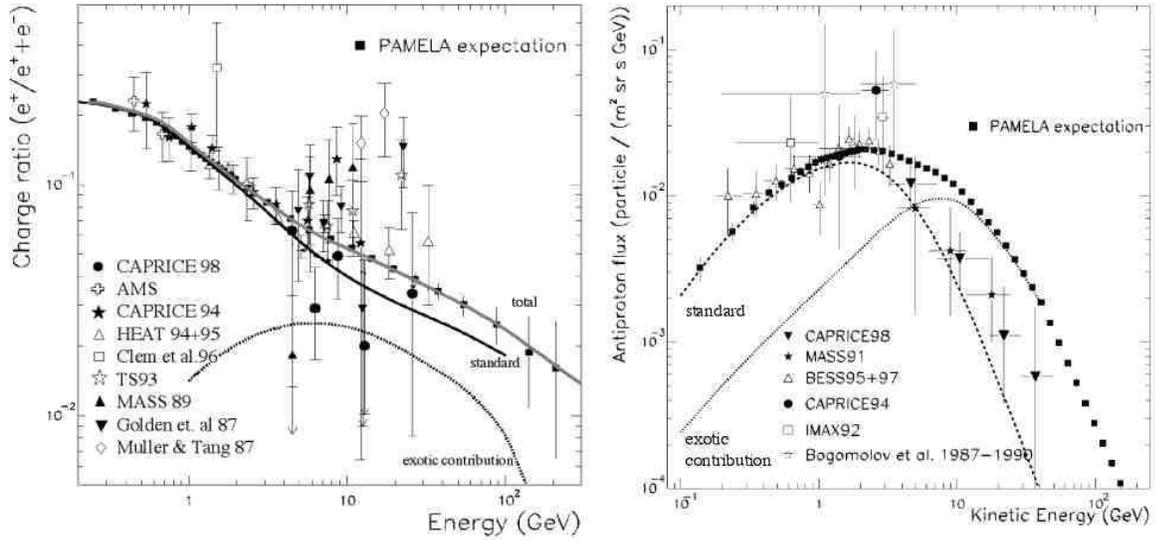,width=1\textwidth,angle=0,clip=}
\caption{\label{pamela2} \it  Distortion of the secondary 
positron fraction (on the left) and  secondary antiproton flux (on the right) induced by a signal from a
heavy neutralino. The PAMELA expectation in the case of  exotic contributions are shown by black squares} 
\end{figure}

As an example, the expected data
from the experiment PAMELA  in the annihilation scenario for three years of operation are shown
by  black squares in figure~\ref{pamela2} for both the positron and antiproton fluxes.

\subsection{  The PAMELA telescope}

The PAMELA telescope, based on the experience gained in the WiZard balloon flights and in the WIZARD\cite{wizard} on ASTROMAG and Mass-Sat \cite{mass_sat} proposals,  is composed by:
\vskip 0.2cm
\par\noindent 1.	A magnetic spectrometer to determine the sign of the electric charge  and to measure the momentum of the particles. The magnetic
system is composed by five permanent magnet of Nd-Fe-B, each 8 cm high, that provide a field inside
the tracking volume of about 0.4 T. There are six planes of silicon micro-strip detectors for tracks reconstruction 
 with  a spatial resolution of 3 $\mu$m in the bending view giving a Maximum Detectable Rigidity (MDR) of 740 GV/c.

\par\noindent	
$\bullet$	A Transition Radiation Detector (TRD) for distinguishing at the level of $10^{-2}$ between electromagnetic  and  hadronic showers  from 1 GeV to about 1 TeV;  the TRD is based on small diameter straw tubes filled with Xe-CO$_2$ mixtures and arranged in
double layer planes interleaved by carbon fiber radiator. 

$\bullet$ A $16 X_0 $ silicon imaging calorimeter to discriminate  at the level of $10^{-4} - 10^{-5}$ between electromagnetic  and  hadronic showers.  It is composed by  silicon strip sensors, interleaved with tungsten plates as converters. 
The energy resolution for positrons is of the order of 15\% / E$^{1/2}$.
In self triggering mode the calorimeter  geometric factor is about 470 cm$^2 sr$, allowing the measuring with useful statistics of the electron flux up to 2 TeV.
 
\par\noindent	
$\bullet$	Six scintillation counter (each 7 mm thick) hodoscopes  for the event trigger
and  for TOF measurements with a time resolution of $\sim$ 110 ps to provide  albedo discrimination and particle identification up to 3 GeV/c.

\par\noindent	
$\bullet$  An additional plastic scintillator  (S4 in the  figure\ref{pamela2} ) and a neutron detector composed by 36 $^3$He counters in a polyetilen moderator allow, together with the imaging calorimeter,  to extend the energy range for primary protons and electrons up to 10 TeV. 

\par\noindent	
$\bullet$ A set of scintillation counters,  covering part of the top edge, the lateral sides of the magnetic
spectrometer and the bottom part of the calorimeter, completes the telescope, for a further labeling
of contaminating events.

Others characteristics of the PAMELA instrument are:
	
\par\noindent 1.	an acceptance (geometrical factor) of 20.5 cm$^2$ sr; 

\par\noindent 2.	a total volume of  $120{\rm x} 40{\rm x}45$ cm$^3$;

\par\noindent 3.	a total mass of 470 kg ;

\par\noindent 4.	a power consumption of 360 W.

The PAMELA instrument will be installed onboard  the russian RESURS-DK1 satellite, built by TsSKB-Progress.
It will be launched in 2003 from Baikonur with a Soyuz TM rocket and placed on an elliptic orbit at altitude 300-600 Km 
and an inclination of 70.4$^\circ$ for a mission at leas three years long. 

 Averaging on the solar activity, in the first three years of the PAMELA flight
(2003-2006) we expect to collect  the following approximate numbers for particles,
antiparticles and some nuclei:

\begin{table} [h]         
\begin{center}
\begin{tabular}{|l l|l l|}\hline 
	protons	   &  3 $ 10^8 $   & anti-protons &  3 $ 10^4 $ \\ \hline
	electrons	 &  3 $ 10^6 $   & positrons    & 	1 $ 10^5 $ \\ \hline
	He nuclei	 & 	4 $ 10^7 $   & Be nuclei    & 	4 $ 10^4 $ \\ \hline
	C nuclei	  &  4 $ 10^5 $   & anti-nuclei limit &  6 $ 10^{-8}$  (90\% C.L.)\\ \hline
\end{tabular}
\end{center}
\end{table}

%

\end{document}